\documentclass[12pt]{article}

\usepackage[top=1in, left=1in, right=1in,
bottom=1in]{geometry}  
\usepackage[parfill]{parskip}% Activate to begin paragraphs with an empty line rather than an indent
\usepackage{graphicx}
\usepackage{amssymb}
\usepackage{epstopdf}
\usepackage{bbm}
\usepackage{amssymb}
\usepackage{amsmath}
\usepackage{amsfonts}

\usepackage{lscape}
\usepackage{rotating}
\usepackage{setspace}
\usepackage{threeparttable}
\usepackage{booktabs}
\usepackage[compact]{titlesec}
\usepackage[table]{xcolor}
\usepackage{appendix}

\usepackage{lmodern}
\fontfamily{lmtt}\selectfont
\usepackage[T1]{fontenc}

\usepackage{natbib}

\bibpunct{(}{)}{;}{a}{}{,}

\usepackage{titling}
\newcommand{\subtitle}[1]{%
  \posttitle{%
    \par\end{center}
    \begin{center}\large#1\end{center}
    \vskip0.5em}%
}

\begin{document}
\pagestyle{plain}

\newcommand{\blind}{0}

\newcommand{\tit}{\Large Variable-Ratio Matching with Fine Balance in a Study of Peer Health Exchange}

\if0\blind

{\title{\tit\thanks{For comments and suggestions, we thank Paul Rosenbaum.}}
\author{Luke Keele\thanks{Penn State University, University Park, PA and the American Institutes for Research, Washington D.C., Email: ljk20@psu.edu.} \and
      Sam Pimentel \thanks{University of Pennsylvania, Philadelphia, PA, Email: sdbpimentel@gmail.com} 
\and Frank Yoon\thanks{Mathematica Policy Research, Princeton, NJ, Email: fyoon@mathematica-mpr.com}
      }

\date{\today}

%\date{First draft: March 27, 2011 
%\\ 
%\medskip
%This draft: \today}

\maketitle
}\fi

\if1\blind
\title{\bf \tit}
\maketitle
\fi

\begin{abstract}

In observational studies of treatment effects, matched samples are created so treated and control groups are similar in terms of observable covariates.  Traditionally such matched samples consist of matched pairs. If a pair match fails to make treated and control units sufficiently comparable, alternative forms of matching may be necessary. One general strategy to improve balance is to match a variable number of control units to each treated unit.  A more tailored strategy is to adopt a fine balance constraint. Under a fine balance constraint, a nominal covariate is exactly balanced, but it does not require individually matched treated and control subjects for this variable. In the example, we seek to construct a matched sample for an ongoing evaluation of Peer Health Exchange, an intervention in schools designed to decrease risky health behaviors among youth. We find that an optimal pair match that minimizes distances between pairs creates a matched sample where balance is poor.  Here we propose a method to allow for fine balance constraints when each treated unit is matched to a variable number of control units, which is not currently possible using existing matching algorithms. Our approach uses the entire number to first determine the optimal number of controls for each treated unit. For each strata of matched treated units, we can then apply a fine balance constraint. We then demonstrate that a matched sample for the evaluation of the Peer Health Exchange based on a variable number of controls and fine balance constraint is superior to simply using a variable-ratio match.

\end{abstract}

%\vspace*{.3in}

\begin{center}
\noindent\textsc{Keywords}: 
%\small 
{Matching; Fine Balance; Observational Study; Optimal Matching; Entire Number}
%\normalsize
\end{center}
\clearpage
\doublespacing

\section{Introduction}

\subsection{A Motivating Example: Peer Health Exchange}

Many under-resourced high schools lack any curriculum on health education.  Health education courses cover such topics as sexual health, substance abuse, and instruction on nutrition and physical fitness. Peer Health Exchange (PHE) is a nonprofit organization established in 2003 that seeks to provide health education in underprivileged high schools that lack such a curriculum \citep{Sloane:1993, White:2009}. Instead of providing curricular materials to schools, the PHE relies on a specific model of health education.  Schools that partner with PHE offer health education through the use of trained college student volunteers.  College student volunteers serve as peer health educators for high or middle school students. Using peer educators to address sensitive topics such as sexual health is thought to allow for a stronger connection between the students and educators.   The PHE model is designed to modify student behaviors and attitudes in the areas of substance abuse (use of alcohol, tobacco, and illicit drugs), sexual health (use of contraception, pregnancy, sexual health risks), and mental health. The effectiveness of the PHE model has not been rigorously tested.  Early research has shown that peer health educators can be more effective than community health nurses \citep{Dunn:1998a,Forrest:2002}. A review of extant research by \citet{Kim:2008} found that peer-led sex education improved knowledge, attitudes, and intentions, but actual sexual health outcomes were not improved. 

As part of a larger multiphase study to evaluate the effectiveness of the PHE model, schools in a large Midwestern city were recruited to implement the PHE model. At the same time a set of comparison schools were selected via a pairwise Mahalanobis match and recruited into the study. Schools were matched on enrollment, percentage of students that were English language learners, percentage of students receiving special education services, the percentage of students eligible to participate in the free or reduced price lunch program, the percentage of African American students, the percentage of Latino students, the average school score on a state standardized test of English and language achievement, and the average school score on a state standardized test of mathematics achievement. Within these schools, students completed a battery of survey items on health behaviors before the PHE curriculum was implemented in the treated schools. 

The PHE curriculum was implemented in the treated schools during the Spring semester of 2014.  Students then received a follow up survey at the end of the school year.  The follow-up survey was composed of the same battery of items on health behaviors administered at baseline.  These measures serve as outcomes in the PHE evaluation. In total 121 students completed the PHE curriculum.  From the control schools, a pool of 357 students were available as controls. Hereafter, we interchangeably refer to treated students as ``PHE'' students.

\subsection{An Initial Pair Match}
\label{ssec:pair}

Even though the schools that did not receive the PHE curriculum were similar to those schools that did, the study design also called for matching students at the individual level. The student-level match was included since differing outcomes among students may reflect initial differences in student-level covariates between the treated and control groups rather than treatment effects \citep{Cochran:1965,Rubin:1974}.   Pretreatment differences or selection biases amongst subjects come in two forms: those that have been accurately measured, which are overt biases, and those that are unmeasured but are suspected to exist which are hidden biases.  Matching methods are frequently used to remove overt biases. Matched samples are constructed by finding close matches to balance pretreatment covariates \citep{Rosenbaum:1983}. Ideally, such matches are constructed using an optimization algorithm \citep{Rosenbaum:1989,Ming:2000,Hansen:2004,Zubizarreta:2012}. 

\begin{figure}[htbp]
\begin{center}
\includegraphics[scale=.9]{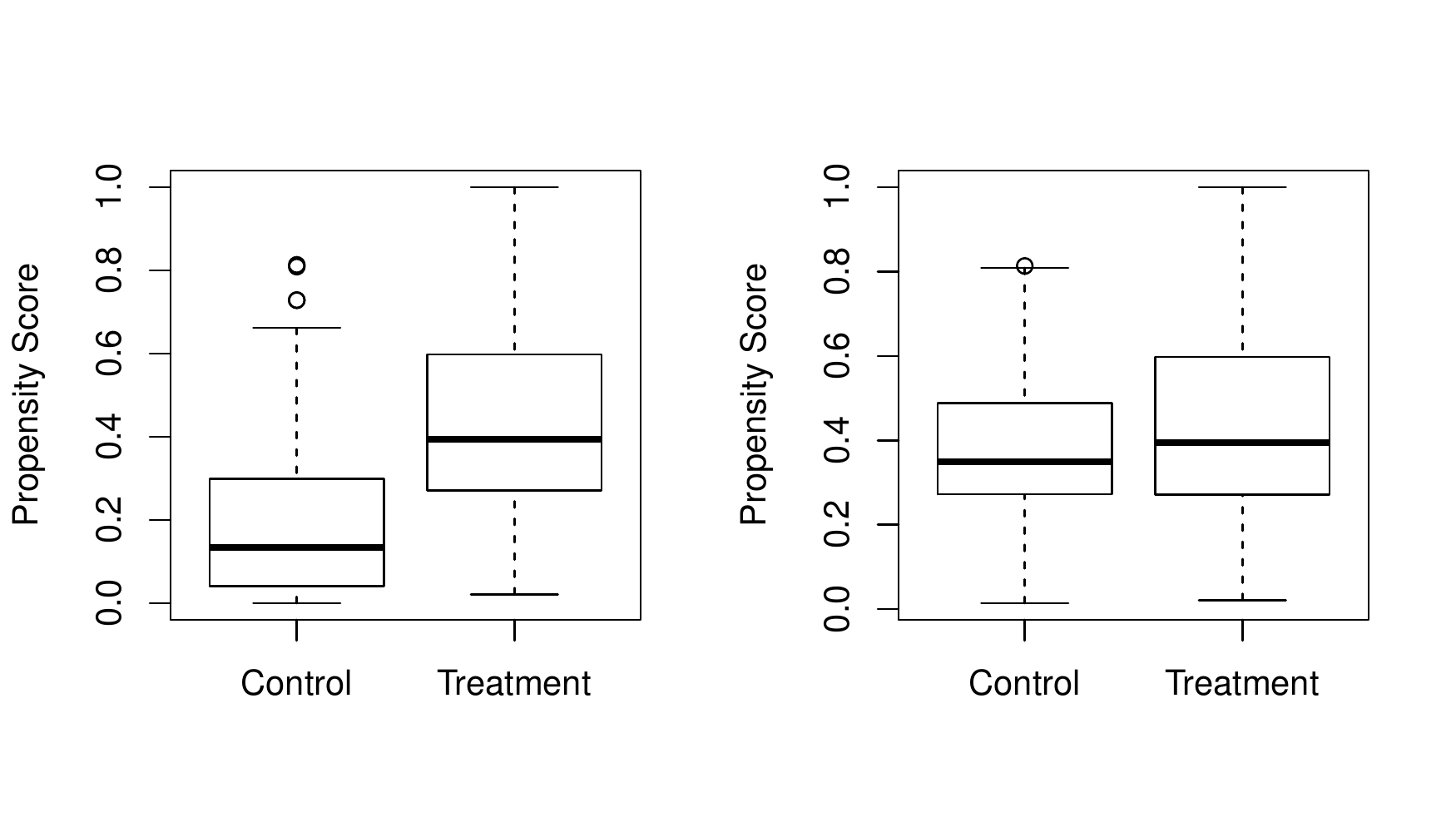}
\end{center}
\vspace{-.4in}
\caption{Boxplots of propensity scores before matching and after pair matching.
\label{fig:pscore}}
\end{figure}

In all, 21 covariates were available describing student demographics and behaviors in four health-related subject areas. In the appendix, we report means for the treated and control groups, standardized differences in means (difference in means divided by the pooled standard deviation between groups before matching), and the $p$-value for the difference in means before any student level matching was implemented. Before matching, the PHE students were much more likely to be African American, less likely to be female, and more likely to be eligible for the free or reduced price lunch program.  The measure of eligibility for the free lunch program is a key covariate as it is the sole indicator of socio-economic status for the students in the study, and it is substantially imbalanced with a standardized difference of 0.616 in the unmatched data. PHE students were also more likely to have a higher incidence of drug use and sexual activity. After the pair match, 15 covariates had imbalances of greater than 0.10 as measured by the standardized difference. We found that 10 covariates also had imbalances that were statistically significant at the 0.05 level or below. The left panel of Figure~\ref{fig:pscore} contains a box plot of the estimated propensity score for the both the treated and control groups. The distribution of the propensity scores for the treated students is shifted much higher than the distribution of propensity scores for student in the comparison group.

To reduce overt bias due to these imbalances, we first implemented a pair match. For this match, we sought to minimize distances based on a rank-based Mahalanobis distance metric. We also applied a caliper to the estimated propensity score through a penalty function. We set the caliper to be 0.5 times the standard deviation of the estimated propensity score. See \citet[ch 8]{Rosenbaum:2010} for an overview of both the distance metric and calipers enforced via penalties. We implemented this match using the \texttt{pairmatch} function in Hansen's \citeyearpar{Hansen:2007} \texttt{optmatch} library in \texttt{R}.  For a number of measures, survey responses were missing. To understand, whether this pattern of missingness differed across the treated and control groups, we use a method recommended by \citet{Rosenbaum:2010}. Under this procedure, we imputed missing values using the mean for that covariate, but we then created a separate indicator for whether the value was missing. We then checked balance on these measures of missingness to understand whether the patterns of missingness were imbalanced across treated and control groups.

\begin{table}[htbp]
\centering
\caption{Covariate balance on pair matched data for PHE evaluation (1/0 indicates binary covariate. Missing values were imputed with means, and missing data indicators denote whether value is missing.)} 
\label{tab:pmatch}
\begin{tabular}{lrrrr}
\toprule
 & Mean C & Mean T & Std Diff. & P-val \\ 
\midrule
\multicolumn{5}{c}{Demographics}		\\
African American 1/0 & 0.843 & 0.802 & -0.087 & 0.318 \\ 
  Multi-Racial 1/0 & 0.008 & 0.033 & 0.207 & 0.180 \\ 
  White 1/0 & 0.008 & 0.008 & 0.000 & 1.000 \\ 
  Hispanic 1/0 & 0.140 & 0.149 & 0.023 & 0.842 \\ 
  Female 1/0 & 0.554 & 0.463 & -0.190 & 0.056 \\ 
  Disability type 1 1/0 & 0.025 & 0.033 & 0.042 & 0.654 \\ 
  Disability type 2 1/0 & 0.008 & 0.017 & 0.074 & 0.564 \\ 
  Disability type 3 1/0 & 0.149 & 0.198 & 0.042 & 0.654 \\ 
  Free or reduced price lunch 1/0 & 0.950 & 0.917 & -0.075 & 0.206 \\

\multicolumn{5}{c}{Substance Abuse	History}		\\
 Marijuana use 1/0 & 0.140 & 0.208 & 0.206 & 0.042 \\ 
  Drunk in past 30 days 1/0 & 0.084 & 0.151 & 0.219 & 0.040 \\ 
  5 or more drinks in past 30 days & 0.008 & 0.034 & 0.189 & 0.164 \\ 
  Drug use past 30 days & 0.223 & 0.322 & 0.260 & 0.008 \\ 
  Type of drugs used & 0.231 & 0.512 & 0.289 & 0.016 \\

\multicolumn{5}{c}{Sexual Behaviors}		\\
  Number of sexual partners & 0.349 & 0.458 & 0.124 & 0.300 \\ 
  Ever had sex 1/0 & 0.265 & 0.349 & 0.218 & 0.022 \\ 
  Understand cause of pregnancy 1/0 & 0.701 & 0.757 & 0.142 & 0.208 \\ 
  Can obtain contraception 1/0 & 0.466 & 0.440 & -0.054 & 0.576 \\ 
  Perception of sex safety & 2.956 & 2.944 & -0.021 & 0.816 \\ 

\multicolumn{5}{c}{Other Items}			\\
  Decision-making skill & 3.092 & 3.048 & -0.091 & 0.302 \\ 
  Knowledge of healthy eating & 0.821 & 0.799 & -0.081 & 0.166 \\ 
  Number of times eating healthy & 2.446 & 2.507 & 0.058 & 0.522 \\ 
  Number of days physically active & 3.909 & 4.129 & 0.097 & 0.250 \\ 

 \multicolumn{5}{c}{Missing Data Indicators}	\\
  Marijuana missing 1/0 & 0.000 & 0.008 & 0.091 & 0.318 \\ 
  Drinking 30 missing 1/0 & 0.008 & 0.025 & 0.148 & 0.158 \\ 
  Drink 5 missing 1/0 & 0.008 & 0.058 & 0.350 & 0.014 \\ 
  Sex partners missing 1/0 & 0.041 & 0.050 & 0.046 & 0.564 \\ 
  Had sex missing 1/0 & 0.041 & 0.050 & 0.046 & 0.564 \\ 
  Pregnancy Missing 1/0 & 0.008 & 0.017 & 0.053 & 0.564 \\ 
  Contraception missing 1/0 & 0.008 & 0.025 & 0.089 & 0.318 \\ 
  Sex safe missing 1/0 & 0.008 & 0.008 & 0.000 & 1.000 \\ 
  Decision-making missing & 0.025 & 0.025 & 0.000 & 1.000 \\ 
  Eating knowledge missing & 0.579 & 0.612 & 0.068 & 0.432 \\ 
  Healthy eating missing 1/0 & 0.000 & 0.000 & 0.000 & 1.000 \\ 
  Active missing 1/0 & 0.000 & 0.017 & 0.148 & 0.158 \\ 

\bottomrule
\end{tabular}
\end{table}

The results from this match are contained in Table~\ref{tab:pmatch} and the right-hand panel of Figure~\ref{fig:pscore} which contains box plots of the estimated propensity scores for the treated and control groups. The results from the pair match were less than satisfactory. Although the pair match is an improvement over the unmatched sample, a number of significant imbalances remained. Of the 21 covariates, 11 still had imbalances in which the standardized difference exceeded 0.10. Within the indicators for missing values, two still displayed imbalances of standardized differences greater than 0.10, and two more were nearly 0.10. A general rule of thumb is produce standardized differences of less than 0.20 and preferably 0.10 \citep{Rosenbaum:2010}. Might a better match be possible?

\subsection{Alternatives to a Pair Match}

When pair matching is unsatisfactory, many alternative matching methods exist.  Broadly one can attempt to improve balance generally or by targeting specific covariates.  Two alternatives to a pair match that may reduce overt bias generally are full matching and matching with variable ratios of control units.  A full match is the most general form of optimal matching \citep{Rosenbaum:1991,Hansen:2004,Rosenbaum:2010}. Under full matching, we create matched sets in which each matched set has either 1 treated unit and a variable number of controls or 1 control unit with a variable number of treated units. A second alternative to pair matching is to match each treated unit with a variable number of control units. Given that the ratio of treated to control students is approximately 1:3  in the PHE evaluation data, matching with a variable treatment:control ratio would be a feasible strategy to improve the quality of the matches. \citet{Ming:2000} demonstrated that variable-ratio matching often removes more overt bias than pair matching.

Alternatively, we might attempt to improve balance on specific covariates. This can be accomplished through the use of calipers, exact matching on covariates, penalty terms, or fine and near-fine balance constraints.  Fine balance constraints may be useful in the PHE evaluation match since we have a large number of discrete covariates that remain imbalanced.  Why might fine balance be useful?  Matching in observational studies balances covariates stochastically, but may not have much success in balancing many small strata on discrete covariates because such imbalances can occur by chance. Fine balance forces exact balance at all levels of a nominal variable but places no restriction on individual matched pairs---any one treated subject can be matched to any one control \citep{Rosenbaum:2007b}.   Fine balance is not always possible, when this occurs near-fine balance is one alternative \citep{Yang:2012}.  Matches with fine and near-fine balance constraints can be computed for a fixed-ratio match via an augmentation of the distance matrix. Although we can improve balance on discrete covariates by requiring exact matches, exact matching tends to reduce the number of possible matches making it more difficult to improve balance on other covariates.

Given the failure of the pair match and the ratio of treated to control units in the PHE evaluation data, one natural design for the PHE evaluation would be to implement a match with a variable treatment:control ratio and use fine or near-fine balance to remove any residual imbalances.  However, no algorithms currently exist that would allow us to use fine balance when matching with a variable number of controls. Below we develop a new algorithm that minimizes total distance among treated and control units via a variable-ratio match, but also allows for fine or near-fine balance constraints.

%\label{sec:review}
%\label{sec:alg}
%\label{ssec:exp}
%\label{ssec:gp}
%\label{sec:phe}
%\label{ssec:comp}
%\label{sec:dis}

\subsection{Outline}

The article is organized as follows. Section~\ref{sec:review} develops notation and reviews variable-ratio matching and fine balance in greater detail.  In particular, we introduce the entire number, a form of design-based variable-ratio matching that will make incorporation of fine balance constraints possible \citep{Yoon:2009}. In Section~\ref{sec:alg} we detail the proposed algorithm.  We first build intuition in Section~\ref{ssec:exp} and then describe the general procedure in Section~\ref{ssec:gp}.  Next, we demonstrate the use of our method with the PHE evaluation data in Section~\ref{sec:phe}.  We then compare this match to two more conventional matches in Section~\ref{ssec:comp}. Section~\ref{sec:dis} concludes.

\section{Review and Definitions of Variable-Ratio Matching, the Entire Number, and Fine Balance}
\label{sec:review}

In the last section, we observed that a pair match was insufficient to remove overt bias in the data on the effectiveness of the PHE intervention. One alternative form of matching is variable-ratio matching.   Next, we review variable-ratio matching via the ``entire number.''

\subsection{Review: Variable-Ratio Matching and the Entire Number}

To fix the concept of a variable-ratio match, we first define some notation. A match consists of $i = 1,\dots,I$ matched sets.  Each matched set $i$ may contain at least $n_i > 2$ subjects indexed by $j = 1,\dots,n_i$.  Within the matched set, we use an indicator $Z_{ij}$ to denote exposure to the PHE treatment, where $Z_{ij} = 1$ if a student attended the PHE program and $Z_{ij} = 0$ if the student does not. Under the most general matching, there are $m_i$ students with $Z_{ij} =1$ and $n_i - m_i = k_i$ students where $Z_{ij} =0$ (with $m_i$ and $k_i>0$). Under variable-ratio matching, we fix $m_i = 1$ within each matched set, and $k_i$ is allowed to vary from matched set to matched set. For each set $i$,  we may also require each treated student to be matched to at least $\alpha \geq 1$ and at most $\beta \geq \alpha$ controls, i.e. $\alpha \leq k_i \leq \beta$. If $\alpha = \beta = 1$, the matched set is a pair match.  If we set $\beta=3$ and $\alpha =1$ each treated unit may be matched to 1,2 or 3 controls.  Often we place an upper limit on $\beta$ since large matched sets reduce efficiency \citep{Hansen:2004}. In general there is little to gain from $\beta = 10$ and generally $\beta=5$ is sufficient. See \citet{Ming:2000} for a more detailed discussion of the gains from larger matched sets. Under variable-ratio matching, the size of $n_i$ is permitted to vary with $i$, and we wish to select the size of each $k_i$ to minimize a distance criterion.  \citet{Ming:2001} presented one algorithm to optimally select the size of each matched set.  One alternative to their algorithm is the entire number \citep{Yoon:2009}. Next, we review the entire number and how it may be used to select optimal $k_i$ values.

We define $\mathbf{x}$ as a matrix of covariates that are thought to be predictive of treatment status, and  $e(\mathbf{x}) = P(Z_{ij}=1|\mathbf{x})$ as the conditional probability of exposure to treatment given observed covariates $\mathbf{x}$. The quantity $e(\mathbf{x})$ is generally known as the propensity score \citep{Rosenbaum:1983}. The entire number $\nu(\mathbf{x})$ is equal to the inverse odds of $e(\mathbf{x})$, i.e. $\nu(\mathbf{x}) = \frac{1 - e(\mathbf{x})}{e(\mathbf{x})}$. It can be used to select optimal values for $k_i$.  Define $\mathbf{x}_t$ as covariate values for treated students, and $\nu(\mathbf{x}_t)$ is the entire number for treated unit $t$. The entire number represents the average number of controls that are available for matching to a treated subject with covariate value $\mathbf{x}_t$ \citep{Yoon:2009}.  There is an intuitive explanation for why the entire number represents the average number of controls available.  If $e(\mathbf{x}_t) = 1/4$, treated unit $t$ should be matched to $k_i = \frac{1 - 1/4}{1/4} = 3$ controls. That is, given covariate value $\mathbf{x}_t$ the expected number of controls with the same $\mathbf{x}$ is equal to $\nu(\mathbf{x})$ or the inverse odds of the propensity score.

To use the entire number within the context of matching, we use the following procedure. Suppose $\hat{\nu}(\mathbf{x}_t)$ is a non-integer; let $\lfloor \hat{\nu}(\mathbf{x}_t) \rfloor$ denote the first integer immediately below $\hat{\nu}(\mathbf{x}_t)$ (the floor) and $\lceil \nu(\mathbf{x}_t) \rceil$ denote the first integer immediately above (the ceiling). For treated subject $t$ with estimated propensity score $\hat{e}(\mathbf{x}_t)$ and estimated entire number $\hat{\nu}(\mathbf{x}_t) = \frac{1-\hat{e}(\mathbf{x}_t)}{\hat{e}(\mathbf{x}_t)}$, $k_i = \max\{1,\min(\lfloor \hat{\nu}(\mathbf{x}_t) \rfloor, \beta)\}$, so that each treated subject was matched to at least one but at most $\beta$ controls; in between, the $k_i$ is determined by $\lfloor \hat{\nu}(\mathbf{x}_t) \rfloor$. \citet{Yoon:2009} showed that a variable-ratio match based on the entire number is optimal in that it will always remove at least as much bias as a pair match and possibly more. One key advantage of using a variable-ratio match based on the entire number is that it will allow us to include both fine and near-fine balance constraints, while the algorithm developed by \citet{Ming:2001} will not.

\subsection{Review: Fine and Near-Fine Balance}

Fine and near-fine balance are constraints on an optimal match that force a nominal covariate to be exactly or nearly exactly balanced \citep{Rosenbaum:2007b,Yang:2012}.  The match can then focus on the balance of other covariates with the knowledge that this nominal variable will be balanced.  We provide a brief review of fine balance and then discuss how near-fine balance may deviate from fine balance.

Assume there is a discrete, nominal variable with $B \geq 2$ levels, $b=1,\dots,B$, with $n_b \geq 0$ treated subjects at each level $b$, where $n_t = \sum n_b$.  Let $\mathcal{B} \subset \mathcal{C}$ be the subset of controls with level $b$ on the nominal variable $B$ where $\mathcal{C} = \mathcal{B}_1 \bigcup \dots \bigcup \mathcal{B}_b$.  A fixed-ratio match with $\kappa$ controls per treated unit is finely balanced if there are $\kappa n_b$ controls with level $b$ of the nominal variable.  Fine balance forces exact balance at all levels of the nominal variable but places no restriction on individual matched pairs-any one treated subject can be matched to any one control. 

Fine balance is not always possible.  Under near-fine balance, the algorithm comes as close to as fine balance possible \citep{Yang:2012}. \citet{Yang:2012} showed that matches with fine and near-fine balance constraints can be computed by running the assignment algorithm with an augmented distance matrix.  In general, fine and near-fine balance is often used to balance a nominal variable with many levels, a rare binary variable or the interaction of several nominal variables. In the PHE study, there are a large number of nominal covariates where fine balance may address imbalances. While we could require exact matches on these nominal covariates, exact matching tends to restrict the possible matches on other covariates. With fine or near-fine balance, we achieve near or near exact balance on the covariate but we do not place any restriction on individual matches.

\section{Variable-Ratio Matching with Fine and Near-Fine Balance}
\label{sec:alg}

As we highlighted above, variable-ratio matching will often produce substantial gains in bias reduction over pair matching.  Fine and near-fine balance constraints are often useful for balancing the distributions of discrete covariates. One can easily imagine that in the PHE data, it may be desirable to use fine or near-fine balance constraints in combination with variable-ratio matching. However, fine balance is difficult to define and implement within the framework of the typical algorithm used for variable-ratio matching. \citet{Ming:2001} presented an optimal variable-ratio matching algorithm based on a network optimization. When analyzing a variable-ratio match, weights of some kind must be used.  The usual practice is to weight each of the strata (containing one treated unit apiece) so that controls in matched sets with many controls receive low weight and those in matched sets with few controls receive high weight \citep{Rosenbaum:2010}. This varied weighting of control observations is used since a balancing algorithm that treats all controls equally may not produce a covariate distribution similar to the treated population in the re-weighted control population actually used for analysis. 

%Any useful balancing algorithm must account for variable weighting among the controls, but this is not a straightforward task.  
In commonly-used network algorithms for variable-ratio matching, the control ratio for each treated unit is not known in advance but is determined from the data as the match is computed, so the weights for each control unit are not known a priori.  Network formulations for matching with fine balance, however, require all controls in a given matching problem to be weighted equally and depend on a priori knowledge of the values for these equal weights. In order to conduct variable-ratio matching with fine balance, we need an algorithm that defines control ratios for each treated unit in advance so match within groups in which controls have equal, known weights.  The entire number provides a principled and asymptotically optimal way to stratify the population into groups with fixed control-to-treated-unit ratios before matching, and is therefore a natural component of a fine balance algorithm for variable-ratio matching.

\subsection{A Small Example}
\label{ssec:exp}

Consider the following example, in which there are 25 students, 8 of whom received treatment.  The students have entire numbers ranging from 1 to 3, and 12 of them have used drugs in the past 30 days. The goal is to implement a variable-ratio match based on the entire number and enforce a fine balance constraint on the indicator for past drug use. Table~\ref{tab:sexp} summarizes the data. In addition to the covariates shown in Table~\ref{tab:sexp}, we have pairwise distances between treated and control units with the same entire number (perhaps derived from other covariates not provided in the table). The three distance matrices in Table~\ref{tab:sexp_mat} summarize these distances within the three entire number strata. To conduct variable-ratio matching based on the entire number, we perform an optimal match within each of the three entire number strata. In strata 1, we perform an optimal pair match, within strata 2 we perform an optimal 1:2 match, and within strata 3 we perform an optimal 1:3 match.

The left hand side of Table~\ref{tab:sexp_mat} summaries such a match. Within strata 1, we match t$_1$ to c$_5$, t$_2$ to c$_1$, t$_3$ to c$_3$, and t$_4$ to c$_4$ for a total distance of 4.8. In strata 2, we would match t$_5$ to c$_7$ and c$_11$, t$_6$ to c$_10$ and c$_13$, and t$_7$ to c$_8$ and c$_9$. In strata 3, t$_8$ is matched to c$_14$, c$_16$, and c$_17$. This match produces a standardize difference of 0.16 for the measure of past drug use. To improve balance on past drug use, we impose a fine balance constraint. In strata 1, we select only two controls that indicate the use of drugs in the last thirty days, while the treated group contains 3 such units.  To enforce the fine balance constraint, we now pair t$_1$ to c$_6$ and t$_2$ to c$_5$.  This increases the overall distance from 4.8 to 10.5, but contributes to a smaller distance on the indicator of past drug use.  Within stratum 2, fine balance was achieved in the original entire number match since a single treated unit has an indication of drug use in the past thirty days.  This means we that we require exactly two controls with the same status, and we selected exactly two, c$_9$ and c$_{13}$.  Notice that the entire number match with fine balance cannot balance drug use exactly in every stratum. In stratum 3, the treated unit t$_8$ is not a drug user, thus to achieve fine balance we need three controls that have not used drugs in the past thirty days.  The optimal match selects c$_{14}$, c$_{16}$ and c$_{17}$ as the three controls, two of which have used drugs in the past thirty days.  While fine balance is not possible, we achieve near-fine balance by selection of c$_{15}$ instead of c$_{14}$ as one of the three controls. Thus under near-fine balance for drug use, the standard difference for this covariate drops to -0.08 from 0.16, which indicates a reduction in bias of 50\%.

\begin{table}[ht]
\caption{Covariate information for the students in the small example.}
\label{tab:sexp}
\centering

\begin{tabular}{r|rrr}
  \hline
 & Treatment & Drug Use & Entire Number \\ 
  \hline
t1 & 1 & 0 & 1 \\ 
  c1 & 0 & 0 & 1 \\ 
  c2 & 0 & 1 & 1 \\ 
  t2 & 1 & 1 & 1 \\ 
  c3 & 0 & 0 & 1 \\ 
  c4 & 0 & 0 & 1 \\ 
  t3 & 1 & 1 & 1 \\ 
  t4 & 1 & 1 & 1 \\ 
  c5 & 0 & 1 & 1 \\ 
  c6 & 0 & 1 & 1 \\ 
  c7 & 0 & 0 & 2 \\ 
  c8 & 0 & 0 & 2 \\ 
  c9 & 0 & 1 & 2 \\ 
  c10 & 0 & 0 & 2 \\ 
  t5 & 1 & 1 & 2 \\ 
  c11 & 0 & 1 & 2 \\ 
  c12 & 0 & 1 & 2 \\ 
  t6 & 1 & 0 & 2 \\ 
  t7 & 1 & 0 & 2 \\ 
  c13 & 0 & 0 & 2 \\ 
  c14 & 0 & 1 & 3 \\ 
  t8 & 1 & 0 & 3 \\ 
  c15 & 0 & 0 & 3 \\ 
  c16 & 0 & 1 & 3 \\ 
  c17 & 0 & 0 & 3 \\ 
   \hline
   \end{tabular}
\end{table}

\begin{table}[ht]
\caption{Treated-control distance matrices for each entire number stratum in the small example.  Subjects with drug use in the past 30 days are marked in bold. A near-fine balance constraint is used for drug use in one of the matches. The grey shading indicates matched controls for each treated unit within rows.}
\label{tab:sexp_mat}
\centering
\begin{tabular}{r|rrrrrr|rrrrrr}
  \hline
Entire \# & \multicolumn{6}{c|}{Without near-fine balance} & \multicolumn{6}{c}{With near-fine balance} \\
 1 & c$_1$ & {\bf c$_2$} & c$_3$ & c$_4$ & {\bf c$_5$} & {\bf c$_6$} & c$_1$ & {\bf c$_2$} & c$_3$ & c$_4$ & {\bf c$_5$} & {\bf c$_6$} \\ 
  \hline
t$_1$ & 1.2 & 1.5 & 6.7 & 5.2 & \cellcolor{gray!40.0}{1.2} & 3.4 & 1.2 & 1.5 & 6.7 & 5.2 & 1.2 & \cellcolor{gray!40.0}{3.4} \\ 
 {\bf t$_2$} & \cellcolor{gray!40.0}{1.5} & 4.4 & 10.0 & 0.8 & 5.0 & 7.6&1.5 & 4.4 & 10.0 & 0.8 &  \cellcolor{gray!40.0}{5.0} & 7.6 \\ 
  {\bf t$_3$} & 6.0 & \cellcolor{gray!40.0}{1.5} & 3.2 & 5.4 & 6.0 & 5.4 & 6.0 & \cellcolor{gray!40.0}{1.5} & 3.2 & 5.4 & 6.0 & 5.4 \\ 
  {\bf t$_4$} & 8.2 & 5.6 & 7.1 & \cellcolor{gray!40.0}{0.6} & 8.1 & 7.3& 8.2 & 5.6 & 7.1 & \cellcolor{gray!40.0}{0.6} & 8.1 & 7.3 \\ 
   \hline
\end{tabular}

\vspace{1em}

\begin{tabular}{r|rrrrrrr|rrrrrrr}
  \hline
Entire \# & \multicolumn{7}{c|}{Without near-fine balance} & \multicolumn{7}{c}{With near-fine balance} \\
2 & c$_7$ & c$_8$ & {\bf c$_9$} & c$_{10}$ & {\bf c$_{11}$} & {\bf c$_{12}$} & c$_{13}$ & c$_7$ & c$_8$ & {\bf c$_9$} & c$_{10}$ & {\bf c$_{11}$} & {\bf c$_{12}$} & c$_{13}$\\ 
  \hline
{\bf t$_5$} & \cellcolor{gray!40.0}{1.2} & 9.6 & 2.1 & 3.9 & \cellcolor{gray!40.0}{0.9} & 7.0 & 1.7 & \cellcolor{gray!40.0}{1.2} & 9.6 & 2.1 & 3.9 & \cellcolor{gray!40.0}{0.9} & 7.0 & 1.7\\ 
 t$_6$ & 7.8 & 8.8 & 3.7 & \cellcolor{gray!40.0}{3.7} & 6.7 & 6.9 & \cellcolor{gray!40.0}{2.0} & 7.8 & 8.8 & 3.7 & \cellcolor{gray!40.0}{3.7} & 6.7 & 6.9 & \cellcolor{gray!40.0}{2.0} \\ 
  t$_7$ & 10.0 & \cellcolor{gray!40.0}{0.8} & \cellcolor{gray!40.0}{3.9} & 4.0 & 9.4 & 9.1 & 3.9  & 10.0 & \cellcolor{gray!40.0}{0.8} & \cellcolor{gray!40.0}{3.9} & 4.0 & 9.4 & 9.1 & 3.9 \\ 
   \hline
\end{tabular}

\vspace{1em}

\begin{tabular}{r|rrrr|rrrr}
  \hline
Entire \# & \multicolumn{4}{c|}{Without near-fine balance} & \multicolumn{4}{c}{With near-fine balance} \\
 3 & {\bf c$_{14}$} & c$_{15}$ & {\bf c$_{16}$} & c$_{17}$ & {\bf c$_{14}$} & c$_{15}$ & {\bf c$_{16}$} & c$_{17}$ \\ 
  \hline
t$_8$ & \cellcolor{gray!40.0}{1.4}& 3.4 & \cellcolor{gray!40.0}{0.7} & \cellcolor{gray!40.0}{1.8} & 1.4& \cellcolor{gray!40.0}{3.4} & \cellcolor{gray!40.0}{0.7} & \cellcolor{gray!40.0}{1.8} \\ 
   \hline
   \end{tabular}
\end{table}

\clearpage
\subsection{A General Procedure}
\label{ssec:gp}

The general algorithm for variable-ratio matching with fine or near-fine balance constraints is as follows.  Suppose we have a study with subjects $j  \in \{1, \ldots, N\}$, each receiving either a treatment or control.   Suppose also that we have fit an estimated propensity score $\widehat{e}(\boldsymbol{x})$ to the data and let $\widehat{e}_j$ be the estimated propensity to receive treatment for subject $j$. 
\begin{enumerate}
\item Choose a positive integer $K > 1$ as the maximum number of controls that we will allow to be matched to a single treated subject.
\item Define
\begin{align*}
S_1 &= \left(\frac{1}{3}, 1 \right] \\
S_k &= \left(\frac{1}{k+2},\frac{1}{k+1}\right] \quad \quad \text{for $k \in \{2, \ldots, K-1\}$ where applicable} \\
 S_K &= \left[0, \frac{1}{K+1}\right]
\end{align*}
These sets $S_k$ form a partition of the unit interval.
\item For each $k \in \{1, \ldots, K\}$:
\begin{enumerate}
\item Select all the study subjects with $\widehat{e}_j \in S_k$.  For $k \in \{2, \ldots, K-1\}$, these are exactly the subjects with entire numbers in the interval $[k,k+1)$.  For $k = 1$, these are the subjects with entire numbers in $(0, 2)$, and for $k = K$ they are the subjects with entire numbers in $[K, \infty)$.
\item Conduct $1:k$ fixed-ratio matching with near-fine balance among the selected subjects.  Call the resulting match $M_k$.
\end{enumerate}
\item Return $\bigcup^K_{k=1}M_k$ as the final match.
\end{enumerate}   
Briefly stated, the procedure separates study subjects into strata based on their entire numbers and conducts fixed-ratio matching with near-fine balance within strata, using the appropriate treatment:control ratio suggested by the entire number.

We also propose two refinements to variable-ratio matching based on the entire number.  The first refinement is in response to finite sample constraints that may arise.  As $n \longrightarrow \infty$ for fixed $K$, the properties of the entire number ensure that each stratum will have sufficient controls to conduct fixed ratio matching with the appropriate ratio. Within finite samples, however, it may be the case that within some strata there are not a sufficient number of controls to form the appropriate variable-ratio match.  For example, when  $\hat{e}(\mathbf{x}_t) = 1/5$, units within that entire number strata should have four controls matched to each treated unit.  Within a particular finite sample, we may find that within this entire number strata there are 4 treated units and 15 control units. Thus given finite sample constraints, we do not have enough control units for all four treated units.  When this situation arises, we can simply match to the highest ratio possible.  Thus in this strata, we would perform a match with a fixed 1:3 ratio of treated to control units.

Next, we enhance entire number matching to deal with a lack of common support.  A lack of common support occurs when there are neighborhoods of the covariate space where there are not sufficient numbers of treated and control units to make inferences about the treated. A lack of common support manifests itself in a specific way when matching with the entire number. When $\hat{e}(\mathbf{x}_t) \geq 1/2$, entire matching results in a single strata where all units are pair matched.  Within this strata, it may be the case that the number of treated units exceeds the number of control units due to a lack of common support. Matching based on the entire number makes a lack of common support readily transparent, since a pair match becomes impossible within this strata when the number of treated exceeds the number of controls. A lack of common support is endemic to the estimation of treatment effects with observational data and can arise for any estimator of causal effects \citep{Crump:2009,Rosenbaum:2011}.

When there is a lack of common support with entire matching, we can use extant methods to reduce the sample to the region of common support. \citet{Crump:2009} recommend discarding all units with estimated propensity scores outside a specific range.  In our context, this would amount to eliminating all treated units with propensity scores higher than the maximum propensity score among controls, as well as trimming all control units with propensity scores lower than the minimum propensity score among the treated units.  We could also apply the method of \citet{Traskin:2011} to identify and describe the population with insufficient common support in order to exclude individuals in a more interpretable manner. Alternatively, we could use a special case of optimal subset matching \citep{Rosenbaum:2011}. Under this solution, we relabel the control units as treated units and allow for an optimal pair match given the fine balance constraints.  This will discard the treated units that are least comparable to the controls. As such, the algorithm will discard treated units in an optimal manner, retaining the subset of treated units with the lowest overall covariate dissimilarity among the candidate subsets (subject to fine and near-fine balance constraints). 
Discarding treated units implies that we can only estimate the effect of a treatment on marginal students, that is, students who might or might not receive this treatment.  Such a practice seems unobjectionable when the available data do not represent a natural population.  This is true in the PHE evaluation, since only one cohort of students happened to be exposed to the intervention in the first year.  As such, the study population is not representative of a larger population of students.
 
\section{Implementation with the PHE Evaluation Data}
\label{sec:phe}

We next use this new algorithm with the PHE evaluation data. The algorithm was implemented in \texttt{R} using functions from the \texttt{finebalance} library \citep{Yang:2012} and Hansen's \citeyearpar{Hansen:2007} \texttt{optmatch} library, both of which leverage the RELAX-IV algorithm as implemented in FORTRAN by \citet{Bertsekas:1981}; see \texttt{R} Core Development Team \citeyearpar{R:2007} for a discussion of \texttt{R}. In this match, we matched using near-fine balance within entire number strata.  We used 5 entire number strata, with a maximum of $K = 5$ controls per treated unit (for a discussion of this choice of $K$ see the third paragraph under section 4.1). For this match, we continued to minimize distances based on a rank-based Mahalanobis distance. We also maintained a caliper on the estimated propensity score through a penalty function with the caliper set to be 0.5 times the standard deviation of the estimated propensity score. 

We experimented with several possible fine balance constraints (one of the advantages of matching is that the covariate adjustment step is separate from the testing step, so multiple matches can be examined without introducing multiple testing issues). In this process, we focused specifically on closely balancing the indicator for free or reduced price lunch eligibility, since it is the lone indicator for socio-economic status in the data set. One of the most effective constraints required near-fine balance on an interaction of drug use in the past 30 days and the indicator for eligibility in the free or reduced price lunch program. This interaction variable had four categories, one for those who had free lunch but no drug use, one for those with both free lunch and drug use, one for those with drug use but no free lunch, and one for subjects with neither.  We also utilized both of our proposed refinements to matching based on the entire matching. For the stratum with $K=4$, we reduced the match to a 1:1 match due to an insufficient number of control units. The optimal subset procedure also discarded five treated units due to a lack of common support.  As such this match is based on a somewhat different sample than the pair match. However, the number of treated units excluded is small enough that it is still reasonable to compare the two matches. The match resulted in an effective sample size of 135 matched pairs, with 75 pair matches, 23 matches with a ratio of 1:2, 7 matches with a ratio of 1:3, and 11 matches with a ratio of 1:5. \footnote{We also performed one additional match where we dealt with the lack of common support based on trimming the propensity score. Balance for this match was slightly better, but three more treated units were discarded.}

Table~\ref{tab:fine} contains balance statistics for this match. As we noted in Section~\ref{ssec:pair}, a pair match produced results with a high number of imbalances. A combination of a fine balance constraint and variable-ratio matching is a substantial improvement over the pair match. For this match only 3 covariates have a standardized difference of 0.10 or greater with the largest value being 0.153.  In the pair match, 13 covariates had standardized differences of 0.10 or greater with the largest being 0.35. Note that since we apply a near-fine balance constraint the distribution of the free lunch measure and the use of drugs in the past 30 days is not identical as it would be under a strict fine balance constraint.  

%In general, this match results in matched distributions that are much closer to what would be found in a randomized experiment.
%Experiment with qq-plot of p-values.

\begin{table}[htbp]
\centering
\caption{Balance table for entire number match with up to five controls per treated and near-fine balance within strata on $(\text{free lunch} \,\times \,\text{drug use})$. }
\label{tab:fine}
\begin{tabular}{lrrrr}
\toprule
 & Mean C & Mean T & Std Diff. & P-val \\ 
\midrule
\multicolumn{5}{c}{Demographics}		\\
African American 1/0 & 0.834 & 0.802 & -0.071 & 0.115 \\ 
Multi-Racial 1/0	 & 0.013 & 0.026 & 0.089 & 0.770 \\ 
White 1/0 & 0.007 & 0.009 & 0.006 & 1.000 \\ 
Hispanic 1/0	 & 0.143 & 0.155 & 0.035 & 0.096 \\ 
Female 1/0 & 0.541 & 0.466 & -0.155 & 0.286 \\ 
Disability type 1 1/0 & 0.023 & 0.034 & 0.059 & 0.708 \\ 
Disability type 2 1/0 & 0.010 & 0.009 & -0.015 & 0.663 \\ 
Disability type 3 1/0 & 0.140 & 0.207 & 0.059 & 0.897 \\ 
Free or reduced price lunch 1/0 & 0.918 & 0.914 & -0.011 & 0.114 \\ 
\multicolumn{5}{c}{Substance Abuse	History}		\\   
Marijuana use 1/0 & 0.168 & 0.182 & 0.040 & 0.917 \\ 
Drunk in past 30 days 1/0 & 0.112 & 0.131 & 0.059 & 0.770 \\ 
5 or more drinks in past 30 days 1/0 & 0.012 & 0.035 & 0.153 & 0.844 \\ 
Drug use past 30 days 1/0 & 0.254 & 0.293 & 0.094 & 0.940 \\ 
Type of drugs used & 0.301 & 0.336 & 0.033 & 0.745 \\ 
\multicolumn{5}{c}{Sexual Behaviors}		\\    
Number of sexual partners & 0.402 & 0.467 & 0.064 & 0.677 \\ 
Ever had sex 1/0 & 0.307 & 0.354 & 0.112 & 0.874 \\ 
Understand cause of pregnancy 1/0 & 0.711 & 0.747 & 0.088 & 0.889 \\ 
Can obtain contraception 1/0	 & 0.448 & 0.450 & 0.005 & 0.652 \\ 
Perception of sex safety & 2.921 & 2.947 & 0.044 & 0.810 \\ 
\multicolumn{5}{c}{Other Items}			\\  
Decision-making skills & 3.066 & 3.056 & -0.021 & 0.776 \\ 
Knowledge of healthy eating & 0.810 & 0.796 & -0.054 & 0.917 \\ 
Number of times eating healthy & 2.517 & 2.534 & 0.017 & 0.634 \\ 
Number of days physically active & 3.909 & 4.136 & 0.100 & 0.545 \\ 
 \multicolumn{5}{c}{Missing Data Indicators}	\\
Marijuana missing 1/0 & 0.004 & 0.009 & 0.047 & 1.000 \\ 
Drinking 30 missing 1/0 & 0.009 & 0.017 & 0.067 & 0.527 \\ 
Drink 5 missing 1/0 & 0.009 & 0.017 & 0.048 & 0.527 \\ 
Sex partners missing 1/0 & 0.046 & 0.043 & -0.015 & 0.603 \\ 
Had sex missing 1/0 & 0.046 & 0.043 & -0.015 & 0.603 \\ 
Pregnancy Missing 1/0 & 0.016 & 0.017 & 0.008 & 0.784 \\ 
Contraception missing 1/0 & 0.020 & 0.026 & 0.036 & 0.415 \\ 
Sex safe missing 1/0 & 0.011 & 0.009 & -0.032 & 1.000 \\ 
Decision-making missing 1/0 & 0.029 & 0.026 & -0.020 & 0.057 \\ 
Eating knowledge missing 1/0 & 0.596 & 0.603 & 0.015 & 0.578 \\ 
Healthy eating missing 1/0 & 0.000 & 0.000 & 0.000 & 1.000 \\ 
Active missing 1/0 & 0.009 & 0.009 & 0.000 & 0.424 \\ 
\bottomrule
\end{tabular}
\end{table}

\begin{figure}[htbp]
\begin{center}
\includegraphics[scale=.9]{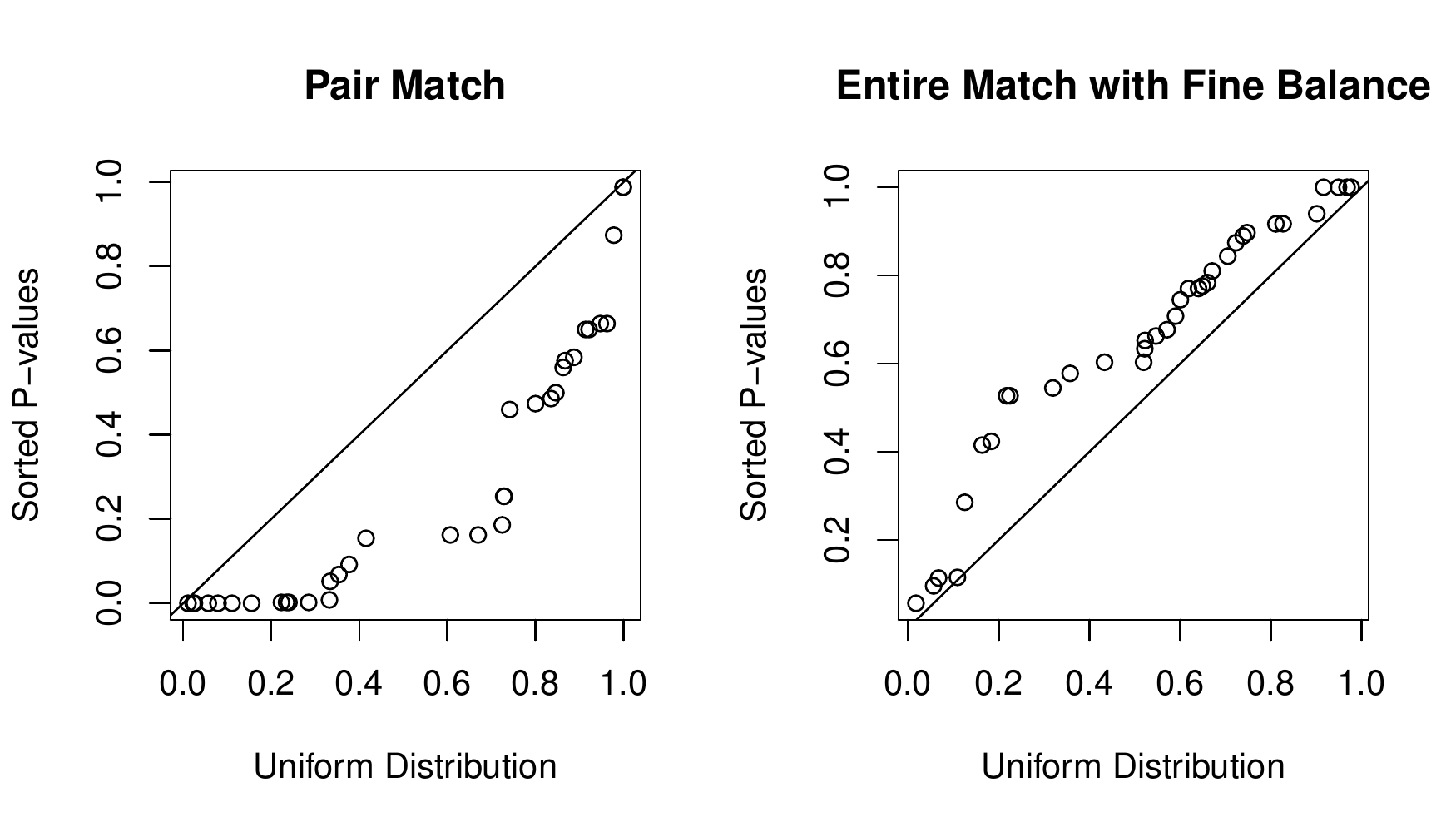}
\end{center}
\vspace{-.2in}
\caption{The quantile-quantile plot compares the 35 two-sample $p$-values with the uniform distribution, with the line of equality.  For the pair match the balance is worse than expected for a completely randomized experiment.  For the variable-ratio match based on the entire number and a fine balance constraint the balance is better than from a randomized experiment.  Balance on observed covariates does not imply balance on unobserved covariates.
\label{fig:pval}}
\end{figure}

In Figure~\ref{fig:pval} we consider one final comparison between the pair match and the match based on our proposed algorithm.  For each match, we compare the two-sample $p$-values for all 35 covariates to the quantiles of the uniform distribution.  For the pair match, the two-sample $p$-values fall below the line of equality which implies that these $p$-values are smaller than they would if from a randomized experiment.  For the variable-ratio match with fine balance, the two-sample $p$-values tend to fall above the  line of equality, which indicates the match produced greater balance than if we had assigned the students to treatment or control at random.  Of course, randomization would also tend to balance unobserved covariates, which matching cannot do.

%%%%%%%%%%%%%%%%%%%%%%%%%%%%%%%%%%%%%%%%%%%
\subsection{Comparison With Two Alternative Matches}
\label{ssec:comp}

We now compare the variable-ratio match with near-fine balance from above to two matches possible with existing algorithms.  We denote these two more conventional matches $\alpha_1$ and $\alpha_2$. Both $\alpha_1$ and $\alpha_2$ are variable-ratio matches based on minimizing the rank-based Mahalanobis distances with a penalty function caliper on the propensity score.  Neither has any fine or near-fine balance constraints. In $\alpha_1$, we implement variable-ratio matching using the entire number.  In $\alpha_2$, we implement a variable-ratio match using the \texttt{optmatch} library \citep{Hansen:2007}. An optimal match with variable controls is also equivalent to solving a particular minimum-cost flow problem in a network \citep{Rosenbaum:1989}. In the statistical package \texttt{R}, the \texttt{fullmatch} function in the \texttt{optmatch} library can be used to create an optimal match with a variable treatment:control ratio based on a distance.  

These additional matches allow us to compare the match in Section~\ref{sec:phe} to two matches that that omit fine balance but use a variable treatment:control ratio.  As we noted variable-ratio matching is a general strategy for reducing bias from imbalances, while fine balance is used to target specific nominal covariates. Specifically under this comparison, we can isolate whether variable-ratio matching reduces bias compared to the pair match in Section~\ref{ssec:pair}. Finally, we can observe a comparison between selecting the ratio of controls via the entire number as opposed to an algorithm based on the minimum cost flow approach.

First, we describe $\alpha_1$.  Initially we used $K = 10$ as the maximum number of controls per treated.  However,  this resulted in many small strata (4 of the 10 strata produced had fewer than 20 subjects total) in which match quality was often poor.  For example, in the stratum with entire number 9 there were exactly 9 controls and one treated subject.  This meant all the controls were included in the match, even though the treated subject had used drugs in the past 30 days and only one of the controls had.  To avoid such imbalances we decided to reduce the value of $K$ from 10 to 5.  The resulting match used 13 fewer controls (202 total instead of 215) but had far fewer small strata and allowed better-quality matches on average within strata.

The $\alpha_1$ match is a major improvement over the pair match.  All of the standardized differences, shown in Table~\ref{tab:entire}, were smaller in magnitude than 0.2 and only five were larger than 0.1.  In addition, the $\alpha_1$ match used 76 more controls more than the pair match, which increased its effective sample size from 121 pairs to 134.5 pairs. With this match, we also removed 5 treated units due to a lack of common support. While the $\alpha_1$ match is a clear improvement over the pair match, the fine balance constraints are useful.  Adding the fine balance constraint both reduces the general magnitude of the standard differences and reduces the number of covariates with a standardized difference above 0.10 from 5 to 3. However, the differences between the pair match and a variable-ratio match underscore that such a match can substantially reduce overt bias due to covariate imbalance.

\begin{table}[htbp]
\caption{Balance table for entire number match with up to five controls per treated without any fine balance.}
\label{tab:entire}
\centering
\begin{tabular}{lrrrr}
  \hline
 & mean.ctrl & mean.treat & sdiff & pval \\ 
  \hline
\multicolumn{5}{c}{Demographics}		\\
African American 1/0	 & 0.843 & 0.802 & -0.091 & 0.275 \\ 
Multi-Racial 1/0 & 0.009 & 0.026 & 0.121 & 0.855 \\ 
White 1/0 & 0.009 & 0.009 & 0.000 & 0.784 \\ 
Hispanic 1/0 & 0.140 & 0.155 & 0.043 & 0.182 \\ 
Female 1/0 & 0.560 & 0.466 & -0.194 & 0.441 \\ 
Disability type 1 1/0 & 0.023 & 0.034 & 0.059 & 0.708 \\ 
Disability type 2 1/0 & 0.009 & 0.009 & 0.000 & 1.000 \\ 
Disability type 3 1/0 & 0.140 & 0.207 & 0.059 & 0.880 \\ 
Free or reduced price lunch 1/0 & 0.948 & 0.914 & -0.087 & 0.305 \\ 
\multicolumn{5}{c}{Substance Abuse	History}		\\

Marijuana use 1/0 & 0.155 & 0.182 & 0.077 & 0.995 \\ 
Drunk in past 30 days 1/0 & 0.105 & 0.131 & 0.080 & 0.973 \\ 
5 or more drinks in past 30 days 1/0 & 0.009 & 0.035 & 0.172 & 0.817 \\ 
Drug use past 30 days 1/0 & 0.242 & 0.293 & 0.123 & 0.799 \\ 
Type of drugs used & 0.274 & 0.336 & 0.057 & 0.877 \\ 
\multicolumn{5}{c}{Sexual Behaviors}		\\  
Number of sexual partners & 0.395 & 0.467 & 0.071 & 0.810 \\ 
Ever had sex 1/0	 & 0.294 & 0.354 & 0.143 & 0.659 \\ 
Understand cause of pregnancy 1/0 & 0.709 & 0.747 & 0.093 & 0.834 \\ 
Can obtain contraception 1/0 & 0.456 & 0.450 & -0.011 & 0.731 \\ 
Perception of sex safety	 & 2.942 & 2.947 & 0.007 & 0.911 \\ 
\multicolumn{5}{c}{Other Items}			\\  
Decision-making skills & 3.083 & 3.056 & -0.055 & 0.398 \\ 
Knowledge of healthy eating & 0.818 & 0.796 & -0.081 & 0.544 \\ 
Number of times eating healthy & 2.468 & 2.534 & 0.064 & 0.588 \\ 
Number of days physically active & 3.910 & 4.136 & 0.099 & 0.627 \\ 
\multicolumn{5}{c}{Missing Data Indicators}	\\
Marijuana missing 1/0	 & 0.004 & 0.009 & 0.047 & 0.480 \\ 
Drinking 30 missing 1/0 & 0.009 & 0.017 & 0.067 & 1.000 \\ 
Drink 5 missing 1/0 & 0.009 & 0.017 & 0.048 & 1.000 \\ 
Sex partners missing 1/0 & 0.046 & 0.043 & -0.015 & 0.572 \\ 
Had sex missing 1/0 & 0.046 & 0.043 & -0.015 & 0.572 \\ 
Pregnancy Missing 1/0 & 0.017 & 0.017 & 0.000 & 0.391 \\ 
Contraception missing 1/0 & 0.018 & 0.026 & 0.046 & 1.000 \\ 
Sex safe missing 1/0 & 0.011 & 0.009 & -0.032 & 1.000 \\ 
Decision-making missing 1/0 & 0.029 & 0.026 & -0.020 & 0.057 \\ 
Eating knowledge missing 1/0 & 0.589 & 0.603 & 0.030 & 0.596 \\ 
Healthy eating missing 1/0 & 0.000 & 0.000 & 0.000 & 1.000 \\ 
Active missing 1/0 & 0.009 & 0.009 & 0.000 & 0.689 \\ 

\bottomrule
\end{tabular}
\end{table}

Next, we describe the $\alpha_2$ match. The $\alpha_2$ match used all 121 treated observations which produced an effective sample size of 160 pairs. That is, variable-ratio matching based on the \texttt{fullmatch} function does not remove observations when there is a lack of common support. Including the additional 5 treated units comes at a considerable cost in terms of balance.  The $\alpha_2$ match is only a modest improvement over the pair match. In the $\alpha_2$ match, 13 covariates still have standardized differences above 0.10 just as in the pair match.  However, in general the magnitude of these imbalances are smaller than in the pair match.  The algorithm selected either matched pairs or matched sets with one treated and five controls.  We then altered the $\alpha_2$ match to allow for up to 10 controls for each treated unit.  This matched produced a much wider variety of matched strata, and balance that is more comparable to the $\alpha_1$ match. Generally, we see that to improve on the pair match and produce the results in Table~\ref{tab:fine} required three separate matching strategies. First, we implemented a variable-ratio match using the entire number. We also enforced fine balance constraints and removed five observations that lacked common support.

\begin{table}[ht]
\centering
\caption{Balance table for variable match with up to five controls per treated without any fine balance.}
\label{tab:opt}
\begin{tabular}{lrrrr}
\toprule
 & Mean C & Mean T & Std Diff. & P-val \\ 
\midrule
\multicolumn{5}{c}{Demographics}	\\

African American 1/0 & 0.663 & 0.813 & 0.317 & 0.000 \\ 
  Multi-Racial 1/0 & 0.010 & 0.029 & 0.156 & 0.208 \\ 
  White 1/0 & 0.119 & 0.010 & -0.337 & 0.000 \\ 
  Hispanic 1/0 & 0.123 & 0.137 & 0.041 & 0.692 \\ 
  Female 1/0 & 0.593 & 0.524 & -0.143 & 0.078 \\ 
  Disability type 1 1/0 & 0.040 & 0.029 & -0.053 & 0.608 \\ 
  Disability type 2 1/0 & 0.012 & 0.017 & 0.037 & 0.738 \\ 
  Disability type 3 1/0 & 0.237 & 0.175 & -0.053 & 0.608 \\ 
  Free or reduced price lunch 1/0 & 0.721 & 0.904 & 0.418 & 0.000 \\ 

\multicolumn{5}{c}{Substance Abuse	History}  \\
 
Marijuana use 1/0 & 0.126 & 0.182 & 0.171 & 0.052 \\ 
  Drunk in past 30 days 1/0 & 0.093 & 0.127 & 0.108 & 0.302 \\ 
  5 or more drinks in past 30 days & 0.011 & 0.026 & 0.111 & 0.352 \\ 
  Drug use past 30 days & 0.200 & 0.289 & 0.234 & 0.000 \\ 
  Type of drugs used & 0.274 & 0.449 & 0.180 & 0.064 \\ 

   \multicolumn{5}{c}{Sexual Behaviors}		\\  
Number of sexual partners & 0.313 & 0.400 & 0.099 & 0.208 \\ 
  Ever had sex 1/0 & 0.236 & 0.313 & 0.200 & 0.014 \\ 
  Understand cause of pregnancy 1/0 & 0.783 & 0.745 & -0.096 & 0.272 \\ 
  Can obtain contraception 1/0 & 0.414 & 0.442 & 0.057 & 0.482 \\ 
  Perception of sex safety & 2.990 & 2.942 & -0.079 & 0.326 \\ 

    \multicolumn{5}{c}{Other Items}			\\
Decision-making skill & 3.075 & 3.071 & -0.008 & 0.916 \\ 
  Knowledge of healthy eating & 0.848 & 0.812 & -0.137 & 0.008 \\ 
  Number of times eating healthy & 2.516 & 2.463 & -0.051 & 0.548 \\ 
  Number of days physically active & 3.895 & 4.064 & 0.074 & 0.354 \\ 

   \multicolumn{5}{c}{Missing Data Indicators}	\\  
Marijuana missing 1/0 & 0.010 & 0.006 & -0.046 & 0.706 \\ 
  Drinking 30 missing 1/0 & 0.015 & 0.019 & 0.037 & 0.778 \\ 
  Drink 5 missing 1/0 & 0.015 & 0.044 & 0.205 & 0.132 \\ 
  Sex partners missing 1/0 & 0.046 & 0.037 & -0.046 & 0.504 \\ 
  Had sex missing 1/0 & 0.046 & 0.037 & -0.046 & 0.504 \\ 
  Pregnancy Missing 1/0 & 0.033 & 0.017 & -0.106 & 0.286 \\ 
  Contraception missing 1/0 & 0.042 & 0.023 & -0.101 & 0.328 \\ 
  Sex safe missing 1/0 & 0.015 & 0.006 & -0.091 & 0.286 \\ 
  Decision-making missing & 0.029 & 0.023 & -0.046 & 0.674 \\ 
  Eating knowledge missing & 0.437 & 0.574 & 0.284 & 0.000 \\ 
  Healthy eating missing 1/0 & 0.006 & -0.000 & -0.136 & 0.318 \\ 
  Active missing 1/0 & 0.017 & 0.012 & -0.037 & 0.706 \\ 
\bottomrule
\end{tabular}
\end{table}

\section{Summary}
\label{sec:dis}

By using the entire number, one can construct a fine or near-finely balanced example with a variable treatment:control ratio.  In the PHE example, we produce near-fine balance on an interaction of two nominal covariates while using a variable treatment:control ratio.  The results from this design strategy removed substantially more bias than an optimal pair match. Additionally our results provide two general lessons. First, variable-ratio matching serves as a good general strategy to reduce bias, which confirms results in the literature \citep{Ming:2000}.  Second, fine balance constraints serve a useful means of targeting remaining imbalances.  Finally, the match based on the entire number is comparable and even superior to existing methods for optimal matching with variable numbers of controls.

\clearpage

\appendix
\appendixpage
\addappheadtotoc

Below Table~\ref{tab:unmatch} summary of the balance statistics for the unmatched data.

\begin{table}[htbp]
\centering
\caption{Covariate balance on unmatched data for PHE evaluation (1/0 indicates binary covariate. Missing values were imputed with means, and missing data indicators denote whether value is missing.)} 
\label{tab:unmatch}
\begin{tabular}{lrrrr}
\toprule
 & Mean C & Mean T & Std Diff. & P-val \\ 
\midrule
\multicolumn{5}{c}{Demographics}		\\
African American 1/0 & 0.580 & 0.802 & 0.470 & 0.000 \\ 
  Multi-Racial 1/0 & 0.008 & 0.033 & 0.206 & 0.052 \\ 
  White 1/0 & 0.160 & 0.008 & -0.472 & 0.000 \\ 
  Hispanic 1/0 & 0.143 & 0.149 & 0.017 & 0.874 \\ 
  Female 1/0 & 0.669 & 0.463 & -0.432 & 0.000 \\ 
  Disability type 1 1/0 & 0.042 & 0.033 & -0.046 & 0.664 \\ 
  Disability type 2 1/0 & 0.011 & 0.017 & 0.048 & 0.650 \\ 
  Disability type 3 1/0 & 0.252 & 0.198 & -0.046 & 0.664 \\ 
  Free or reduced price lunch 1/0 & 0.641 & 0.917 & 0.630 & 0.000 \\ 
\multicolumn{5}{c}{Substance Abuse	History}		\\
Marijuana use 1/0 & 0.096 & 0.208 & 0.342 & 0.002 \\ 
  Drunk in past 30 days 1/0 & 0.096 & 0.151 & 0.178 & 0.092 \\ 
  5 or more drinks in past 30 days & 0.014 & 0.034 & 0.147 & 0.162 \\ 
  Drug use past 30 days & 0.140 & 0.322 & 0.477 & 0.000 \\ 
  Type of drugs used & 0.235 & 0.512 & 0.285 & 0.008 \\

\multicolumn{5}{c}{Sexual Behaviors}		\\
Number of sexual partners & 0.172 & 0.458 & 0.326 & 0.002 \\ 
  Ever had sex 1/0 & 0.154 & 0.349 & 0.503 & 0.000 \\ 
  Understand cause of pregnancy 1/0 & 0.812 & 0.757 & -0.139 & 0.186 \\ 
  Can obtain contraception 1/0 & 0.406 & 0.440 & 0.071 & 0.500 \\ 
  Perception of sex safety & 3.034 & 2.944 & -0.150 & 0.154 \\

\multicolumn{5}{c}{Other Items}			\\
  Decision-making skill & 3.084 & 3.048 & -0.075 & 0.474 \\ 
  Knowledge of healthy eating & 0.882 & 0.799 & -0.319 & 0.002 \\ 
  Number of times eating healthy & 2.447 & 2.507 & 0.057 & 0.584 \\ 
  Number of days physically active & 3.692 & 4.129 & 0.192 & 0.068 \\ 

 \multicolumn{5}{c}{Missing Data Indicators}	\\
  Marijuana missing 1/0 & 0.008 & 0.008 & -0.002 & 0.988 \\ 
  Drinking 30 missing 1/0 & 0.008 & 0.025 & 0.147 & 0.162 \\ 
  Drink 5 missing 1/0 & 0.008 & 0.058 & 0.349 & 0.002 \\ 
  Sex partners missing 1/0 & 0.028 & 0.050 & 0.120 & 0.254 \\ 
  Had sex missing 1/0 & 0.028 & 0.050 & 0.120 & 0.254 \\ 
  Pregnancy Missing 1/0 & 0.028 & 0.017 & -0.073 & 0.486 \\ 
  Contraception missing 1/0 & 0.039 & 0.025 & -0.078 & 0.460 \\ 
  Sex safe missing 1/0 & 0.008 & 0.008 & -0.002 & 0.988 \\ 
  Decision-making missing & 0.017 & 0.025 & 0.059 & 0.576 \\ 
  Eating knowledge missing & 0.364 & 0.612 & 0.511 & 0.000 \\ 
  Healthy eating missing 1/0 & 0.003 & -0.000 & -0.061 & 0.560 \\ 
  Active missing 1/0 & 0.011 & 0.017 & 0.048 & 0.650 \\ 
\bottomrule
\end{tabular}
\end{table}

%%%%%%%%%%%%%%%%%%%%%%%%%%%%%%%%%%%%%%%%%%%

\clearpage
\bibliography{keele_revised2}
\bibliographystyle{asa}

%%%%%%%%%%%%%%%%%%%%%%%%%%%%%%%%%%%%%%%%%%%

\end{document}